%% file: Ga2O3.tex
\documentclass[twocolumn,amsmath,amssymb,superscriptaddress,longbibliography]{revtex4-1}
\newcommand{\Ibadai}{Graduate School of Science and Engineering, Ibaraki University 2-1-1 Bunkyo, Mito, Ibaraki, 310-8512, Japan.}
\newcommand{\Kinken}{Institute for Materials Research, Tohoku University (IMR), 2-1-1 Katahira, Aoba-ku, Sendai 980-8577, Japan}
\newcommand{\IMSS}{Muon Science Laboratory and Condensed Matter Research Center, Institute of Materials Structure Science, High Energy Accelerator Research Organization (KEK-IMSS), Tsukuba, Ibaraki 305-0801, Japan}
\newcommand{\Sokendai}{Department of Materials Structure Science, The Graduate University for Advanced Studies (Sokendai), Tsukuba, Ibaraki 305-0801, Japan}
\newcommand{\MSLTokyo}{Materials and Structures Laboratory, Tokyo Institute of Technology, Yokohama, Kanagawa 226-8503, Japan}
\newcommand{\MCES}{Materials Research Center for Element Strategy, Tokyo Institute of Technology (MCES), Yokohama, Kanagawa 226-8503, Japan}
\newcommand{\NIMS}{National Institute for Materials Science (NIMS), Tsukuba, Ibaraki 305-0044, Japan}

\newcommand{\gao}{$\beta$-Ga$_2$O$_3$} 

\newcommand{\rtwo}{I\hspace{-.1em}I}
\newcommand{\rthree}{I\hspace{-.1em}I\hspace{-.1em}I}

\newcommand{\rfour}{I\hspace{-.1em}V}
\newcommand{\rsix}{V\hspace{-.1em}I}

\usepackage{multirow}
\usepackage{graphicx}
\usepackage[dvipdfmx]{color}
\usepackage{dcolumn}

\usepackage{txfonts}
\usepackage{bm}
\usepackage{ulem}

\usepackage[utf8]{inputenc}
\usepackage[T1]{fontenc}
\usepackage{mathptmx}
\usepackage{etoolbox}
\usepackage[hypertex,colorlinks=true,linkcolor=black,citecolor=blue,filecolor=blue,urlcolor=blue,setpagesize=false,nesting=true]{hyperref}

\makeatletter
\def\@email#1#2{%
 \endgroup
 \patchcmd{\titleblock@produce}
  {\frontmatter@RRAPformat}
  {\frontmatter@RRAPformat{\produce@RRAP{*#1\href{mailto:#2}{#2}}}\frontmatter@RRAPformat}
  {}{}
}%
\makeatother

\begin{document}
\title{Local electronic structure of dilute hydrogen in $\beta$-Ga$_2$O$_3$ probed by muons}

\author{M.~Hiraishi}\thanks{email: masatoshi.hiraishi.pn93@vc.ibaraki.ac.jp}\affiliation{\IMSS}\affiliation{\Ibadai}
\author{H.~Okabe}\affiliation{\IMSS}\affiliation{\Kinken}
\author{A.~Koda}\affiliation{\IMSS}\affiliation{\Sokendai}
\author{R.~Kadono}\thanks{email: ryosuke.kadono@kek.jp}\affiliation{\IMSS}\affiliation{\Sokendai}
\author{T.~Ohsawa}\affiliation{\NIMS}
\author{N.~Ohashi}\affiliation{\NIMS}
\author{K.~Ide}\affiliation{\MSLTokyo}
\author{T.~Kamiya}\affiliation{\MSLTokyo} \affiliation{\MCES}
\author{H.~Hosono}\affiliation{\MCES}

\date{\today}

\begin{abstract}
  The local electronic structure of muons (Mu) as dilute pseudo-hydrogen in single-crystalline \gao\ has been studied by the muon spin rotation/relaxation ($\mu$SR). High-precision measurements over a long time range of $\sim$25 $\mu$s have clearly identified two distinct Mu states: a quasi-static Mu (Mu$_1$) and fast-moving Mu (Mu$_2$). By comparing this result with predictions from the recently established ambipolarity model, these two states are respectively attributed to the relaxed-excited states associated with the donor ($E^{+/0}$) and acceptor ($E^{-/0}$) levels predicted by density functional theory (DFT) calculations for the interstitial H. Furthermore, the local electronic structure of Mu$_1$ is found to be an OMu-bonded state with three-coordinated oxygen. The structure is almost identical with the thermal equilibrium state of H, and it is found to function as an electron donor. The other Mu$_2$ is considered to be in the hydride state (Mu$^-$) from the ambipolarity model, suggesting that it is in fast diffusion motion through the short-lived neutral state due to the charge exchange reaction with conduction electrons (Mu$^0+e^-\rightleftarrows$Mu$^-$).
\end{abstract}
\maketitle

\section{Introduction}
Gallium trioxide (\gao) is attracting attention as a material for high-voltage power devices and other applications because of its large band gap ($E_\mathrm{g}\sim$4.9~eV) and associated high critical electric field~\cite{Galazka:18}, where control of electrical activity by $p$/$n$-type carrier doping is one of the critical issues. While Sn or Si doping is known to induce $n$-type conductivity~\cite{Sn_dope_Ueda:97,Orita:00}, it has been pointed out that impurity hydrogen (H) is the possible cause of unintentional carrier doping in as-grown samples~\cite{Galazka:20}. 
\par
From infrared spectroscopic studies of \gao\ containing macroscopic amounts of H, a variety of H states including O-H defects have been observed~\cite{Philip:18_IR1,Qin:19_IR2}. However, their relationship to carrier doping is not always clear due to the lack of information on their local electronic structures and H valence. Furthermore, since isolated H exists only in trace amounts in solids, experimental means to study its microscopic electronic state are limited. In this regard, theoretical studies based on density functional theory (DFT) calculations have played an important role, and for H in \gao\, DFT calculations have shown that interstitial H may be indeed the origin of $n$-type doping~\cite{DFT_Varley:10,DFT_Varley:11,DFT_Li:14}. This prompted us to introduce muons as pseudo-H into \gao\ for investigating the corresponding interstitial H states by analyzing their electronic and dynamical properties in detail by muon spin rotation ($\mu$SR).
\par
The positive muon ($\mu^+$) is an elementary particle with a mass of $0.1126m_\mathrm{p}$ (where $m_\mathrm{p}$ is the proton mass). The local electronic structure of muon in matter is determined by the muon-electron interaction which is practically equivalent to H, as can be seen from the fact that the difference in reduced electron mass between a muon-electron bound state (muonium) and a neutral H atom is only 0.43\%. Therefore, Mu in matter can be regarded as a light isotope of H ($ = ^{0.1126}$H; we introduce the elemental symbol Mu below to denote muon as pseudo-H).
\par
On the other hand, when interpreting the results of the $\mu$SR experiment based on the results of the DFT calculations, there is an important caveat that should be pointed out at this stage.
Generally, muons are implanted as an ion beam of relatively high-energy (typically $\sim$4 MeV), and the associated electronic excitations in the insulator crystals produce free carriers and excitons ($\sim$10$^3$ per muon) \cite{Thompson:74,Alig:75,Itoh:97}. These often propagate rapidly in the crystal, and there is experimental evidence that Mu acts as a capture center for them \cite{Prokscha:07}. Consequently, the electronic state of Mu is a relaxed-excited state generated by the interaction with free carriers and excitons, which can be different from the electronic state under thermal equilibrium as expected from the thermodynamic double charge conversion level ($E^{+/-}$) obtained by DFT calculations.
\par
Another well-known evidence that Mu is in a relaxed-excited state is that two or more different electronic states of Mu are often observed simultaneously in the same material \cite{Lichti:08}. In order to understand the origin of this phenomenon, we have recently conducted an extensive investigation to explore the regularities between the Mu valence states observed in various oxides and the results of previous DFT calculations. As a result, assuming that these relaxed-excited states are associated with acceptor ($E^{0/-}$) and donor ($E^{+/0}$) levels that are predicted by DFT calculations as metastable states, we found that the observed Mu states can be coherently explained by the relationship between acceptor/donor levels and band structure~\cite{Hiraishi:22}.  We call it the ``ambipolarity model'' since such behavior of Mu is a manifestation of the ambipolarity of H through the relaxed-excited states.
\par
In this study, we show that Mu in single-crystalline \gao\ exhibits two different electronic structures corresponding to the relaxed-excited states respectively associated with the donor and acceptor levels in the ambipolarity model. One is in the OMu-bonded state (Mu$_1^+$) corresponding to H serving as a donor, and another is the hydride-like state in rapid motion (Mu$_2^-$).    In particular, the Mu$_2$ state is a component  overlooked in the previous $\mu$SR experiments \cite{musr_King:10,musr_Celebi:12}, and it is speculated to be a transient state diffusing rapidly along the $\langle010\rangle$ axis while undergoing the charge exchange reaction, Mu$_2^-\rightleftarrows{\rm Mu}_2^0+e^-$. The occurrence of charge exchange is supported by the observation that the temperature dependence of the fractional yield of Mu$_2$ exhibits a strong correlation with that of the bulk carrier electron mobility and density. This also implies that the interstitial H can take hydride state under electronic excitation and may exhibit fast diffusion motion, depending on the bulk electronic properties of the host.

 \section{EXPERIMENTAL METHODS and DFT CALCULATIONS}
The sample used in the $\mu$SR experiment was a slab of single crystal (sc-Ga$_2$O$_3$, 10$\times$15$\times$0.6 mm$^3$) with $\langle001\rangle$ plane synthesized by the edge-defined film-fed (EFG) method (provided by Novel Crystal Technology, Inc.)~\cite{Kuramata:16}.
It is reported to have no twin boundary and the lowest carrier density ($N_\mathrm{e}\sim2\times$10$^{17}$cm$^{-3}$) commercially available~\cite{Kuramata:16}.
The electrical conductivity and Hall effect measurements were performed using the PPMS (Quantum Design Co.). Ohmic contacts were formed from vacuum-deposited Ti at room temperature~\cite{Ne_Villora:08,Irmscher11_raw} and gold paste (Seishin Trading Co.~LTD., No.~8556).
The impurity H content in the sc-Ga$_2$O$_3$ sample was estimated to be 3.5$\times10^{18}\mbox{cm}^{-3}$ by thermal desorption spectrometry, which is sufficient to explain the above described $N_\mathrm{e}$.
Details are described in Supplemental Material (SM)~\cite{sm}.
\par
In this study, $\mu$SR measurements and data analysis were also performed on a powder sample (99.99\%, provided by Rare Metallics Co.)~for comparison, and the results were found to be significantly different from those for single crystal. However, these results are excluded from the discussion in this paper, because it is difficult to measure the bulk electronic properties of powder samples which is necessary to consider the cause of the difference.  Instead, they are presented with a brief interpretation in SM \cite{sm}.
\par
Conventional $\mu$SR measurements were performed using the S1 instrument (ARTEMIS) at the Materials and Life-science Experiment Facility, J-PARC  \cite{ARTEMIS}, where high-precision measurements over a long time range of 20--25 $\mu$s can be routinely performed using a high-flux pulsed muon beam ($\sim$$3\times10^4$ $\mu^+$/cm$^2$/s for the single-pulse mode at a proton beam power of 0.8 MW).
The $\mu$SR spectra [the time-dependent decay-positron asymmetry, $A(t)$] which reflects the magnetic field distribution at the Mu site, was measured from room temperature to 4 K under zero field (ZF), weak longitudinal field (LF, parallel to the initial Mu polarization ${\bm P}_\mu$), and weak transverse field (TF, perpendicular to ${\bm P}_\mu$), and were analyzed by least-squares curve fitting~\cite{musrfit}. The background contribution from muons which missed the sample was estimated from $\mu$SR measurements on a holmium plate of the same geometry and subtracted from the asymmetry.
\par
Since the $\mu$SR spectra were found to be dominated by signals from the {\sl diamagnetic} Mu (i.e., Mu$^+$ or Mu$^-$), the data under ZF/LF conditions were analyzed using the dynamical Kubo-Toyabe (KT) function, $G_z^\mathrm{KT}(t;\Delta,\nu,B_\mathrm{LF})$, where $\Delta$ is the linewidth determined by random local fields from nuclear magnetic moments, $B_\mathrm{LF}$ is the magnitude of LF, and $\nu$ is the fluctuation rate of $\Delta$~\cite{KT}.
The KT function is expressed analytically in the case of static ($\nu=0$) and ZF ($B_\mathrm{LF}=0$) conditions,
\begin{equation}
G_z^\mathrm{KT}(t;\Delta,0,0)=\frac{1}{3}+\frac{2}{3}\left(1-\Delta^2t^2\right)e^{-\frac{1}{2}\Delta^2t^2},\label{KT}
\end{equation}
The magnitude of $\Delta$ for a given Mu site is evaluated  by calculating the second moments of dipolar fields from nuclear magnetic moments by the following equation,
\begin{align}
    \Delta^2&\simeq\gamma_\mu^2\sum_{m}f_m\sum_j\sum_{\alpha=x,y}\sum_{\beta=x,y,z}\gamma_m^2({\bf \hat{A}}_{mj}{\bf I}_m)^2\sin^2\Theta_j \label{Eq:dlts}\\
  {\bf \hat{A}}_{mj}&=A^{\alpha\beta}_{mj}=(3r_{mj}^\alpha r_{mj}^\beta-\delta_{\alpha\beta}r_{mj}^2)/r_{mj}^5\quad(\alpha, \beta=x,y,z)\nonumber
\end{align}
where $\gamma_\mu/2\pi=135.539$~[MHz/T] is the muon gyromagnetic ratio, ${\bm r}_{mj}=(x_{mj}, y_{mj}, z_{mj})$ is the position vector from the muon site to the $j$-th nucleus, ${\bm \mu}_m=\gamma_m{\bm I}_m$ is the nuclear magnetic moment of the atom with the natural abundance of $f_m$, $\Theta_j$ is the polar angle of ${\bm r}_{mj}$~\cite{KT}.
Since $^{69}$Ga with $f_1=0.604$ and $^{71}$Ga with $f_2=0.396$ have nuclear spins of $I_m=3/2$, respectively, they have electric quadrupole moments. 
In this case, ${\bm \mu}_m$ is subject to electric quadrupolar interactions with the electric field gradient generated by the point charge of the diamagnetic Mu, leading to the reduction of the effective ${\bm \mu}_m$ (by a factor $\sin^2\Theta_j$) and to the modification of $\Delta$ that depends on the initial muon spin direction relative to the crystal axis. (In the powder sample, the spatial averaging leads to $\langle\sin^2\Theta_j\rangle=2/3$.) The $\Delta$ of the candidate Mu sites inferred from DFT calculations were evaluated using the Dipelec code~\cite{dipelec} with Eq.~(\ref{Eq:dlts}) implemented.

DFT calculations were performed to investigate in detail the local structure of H-related defects using OpenMX~\cite{OpenMX_Ozaki:03} and VASP codes~\cite{VASP}.  Structural relaxation calculations with H (to mimic Mu) using the GGA-PBE exchange correlation function were performed on a $1\times3\times2$ superlattice with cutoff energies of 200~Ry for OpenMX (520~eV for VASP), and $K$ points were set to $3\times4\times3$.  Structures were relaxed until the the maximum force on each atom was less than $3\times10^{-4}$~(Hartree/Bohr) for OpenMX (0.01~eV/\AA\ for VASP).
Additional structural relaxation calculations using the Heyd-Scuseria-Ernzerhof (HSE06) hybrid functional implemented in VASP were performed with $K$ points of $2\times3\times2$ to confirm the structure of the minimum formation energy~\cite{HSE06}.
The Hartree-Fock mixing parameter was set to 0.35, which reproduces the experimental value of the band gap~\cite{DFT_Varley:10,DFT_Varley:11}.

\section{RESULTS}
First, we investigated the initial asymmetry [$A(0)$] of the $\mu$SR spectrum under TF = 2~mT and found that it is independent of temperature ($T$) and nearly constant within experimental error over the entire measured temperature range $4\le T\le300$~K [see Fig.~S5(a) in SM~\cite{sm}].
Since the absence of a paramagnetic Mu at 300 K is confirmed from the TF data, the $T$-independent $A(0)$ values indicates that the muon is mostly in diamagnetic state(s) at all temperatures. (For the possibility of Mu$^0$ with  extremely small hyperfine parameters to exist, see Section \ref{Dcn}.)

\par
\begin{figure}[b]
  \centering
	\includegraphics[width=0.95\linewidth,clip]{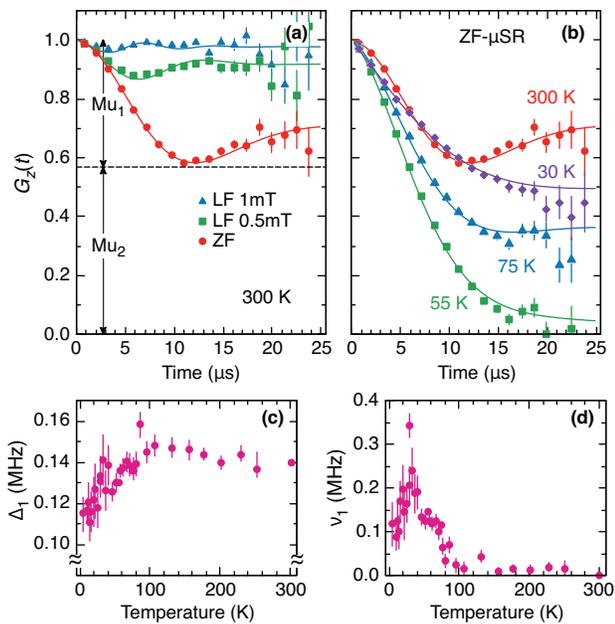}
	\caption{
	(a) ZF and LF $\mu$SR time spectra observed at 300~K ($4.0\times10^8$ positron events collected for ZF and $2.0\times10^8$ for LF), which consists of two components (Mu$_1$ and Mu$_2$). (b) ZF spectra at typical temperatures ($3.0\times10^8$ positron events collected). The solid curves represent the least-square fit by Eq.~(\ref{Eq:ana}). The horizontal dashed line shows $f_2$.  (c), (d) Temperature dependence of the linewidth $\Delta_1$ and fluctuation rate $\nu_1$ for the Mu$_1$ component.}
	\label{fig:asyTdep_par}
\end{figure}

The normalized $\mu$SR time spectra [$G_z(t)\equiv A(t)/A(0)$] at 300~K under ZF and LF ($\parallel \langle001\rangle$) are shown in Fig.~\ref{fig:asyTdep_par}(a). The ZF-$\mu$SR spectrum shows that it consists of two components, one we call Mu$_1$, which partially recovers at $t\gtrsim12$~$\mu\mathrm{s}$ following slow Gaussian relaxation, and Mu$_2$, which exhibits almost no relaxation.  Note that the recovery of $G_z(t)$ characteristic to the 1/3-tail of the Kubo-Toyabe function in Eq.~(\ref{KT}) is clearly visible only for $t\gtrsim$15 $\mu$s. This could not have been detectable by previous $\mu$SR measurements under TF alone \cite{musr_King:10} or those using a continuous muon beam \cite{musr_Celebi:12}, and has been first revealed by precise ZF-$\mu$SR measurements using the high-flux pulsed muon beam at J-PARC.
The suppression of the Gaussian relaxation by the weak LF indicates that the relaxation for the Mu$_1$ component is induced by the quasi-static random local fields from Ga nuclear magnetic moments.
\par
\begin{figure}[b]
  \centering
	\includegraphics[width=0.95\linewidth,clip]{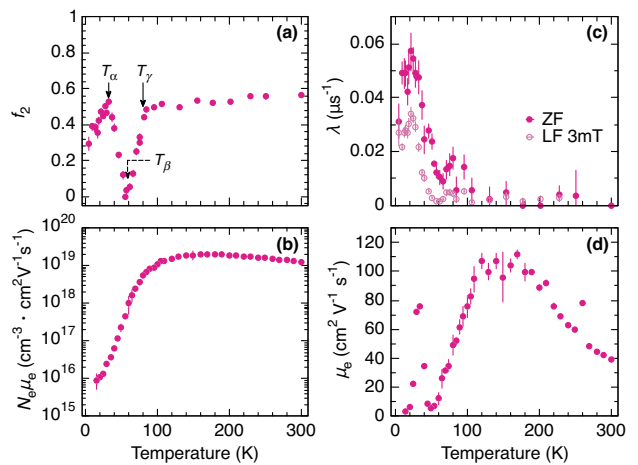}
	\caption{Temperature dependence of (a) fractional yield for the Mu$_2$ component [$f_2\equiv A_2/A(0)$] which is characterized by the changes at $T_\alpha\simeq30$ K, $T_\beta\simeq55$ K and $T_\gamma\simeq80$ K, (b) bulk carrier density ($N_\mathrm{e}$) multiplied by mobility [$\mu_{\rm e}$, shown in (d)], (c) exponential relaxation rate ($\lambda$) at ZF and LF = 3 mT. Note that the vertical axis of (b) is on a logarithmic scale. }
	\label{fig:f2_Ne_lmd}
\end{figure}

It is clear from Fig.~\ref{fig:asyTdep_par}(b) that the relative yield of Mu$_1$ vs Mu$_2$ apparently depends on $T$ at lower temperatures.
Furthermore, the lineshape exhibits change from a Gaussian to exponential-like behavior at low temperatures [see $G_z(t)$ for $t\lesssim 6$ $\mu$s at 30 K].
Considering these features, we analyzed the ZF and LF spectra by global curve-fits using the following function,
\begin{equation}
G_z(t)=f_1G_z^\mathrm{KT}(t;\Delta_1,\nu_1,B_\mathrm{LF})e^{-\lambda t}+f_2,\label{Eq:ana}
\end{equation}
where $f_i$ ($i=1,2$) are the relative yields of the Mu$_i$ components.  The exponential damping with the relaxation rate $\lambda$ is introduced to describe the possible influence of the fluctuating magnetic fields from unpaired electrons (including excited carriers and excitons localized nearby Mu).  As shown by the solid lines in Figs.~\ref{fig:asyTdep_par}(a) (b), curve fit using Eq.~(\ref{Eq:ana}) provides reasonable agreement with the data, and $\Delta_1=0.140(1)$ MHz is obtained from the fit at 300 K. The $T$ dependencies of $\Delta_1$ and $\nu_1$ obtained from the fit are shown in Figs.~\ref{fig:asyTdep_par}(c) and (d), and those of  $f_2$ and $\lambda$ in Figs.~\ref{fig:f2_Ne_lmd}(a) and (c), respectively. 
\par

Above $T_\gamma\simeq80$ K where the $f_1$ ($=1-f_2$) is nealy independent of $T$, $\Delta_1$, $\nu_1$ and $\lambda$ also show similar trends; $\Delta_1$ increases slightly to $\sim$0.15 MHz as $T$ decreases, while $\nu_1$ and $\lambda$ shows a constant value close to zero ($\ll\Delta_1$).  This indicates that the Mu$_1$ component is quasistatic for $T\gtrsim T_\gamma$. Meanwhile, the rapid motion of Mu$_2$ is inferred from the fact that there is no interstitial sites free of local magnetic fields from Ga nuclear magnetic moments (with 100\% natural abundance); $\Delta$ calculated at any site is always larger than $\sim$0.15 MHz for the unrelaxed lattice, which is only compatible with the situation where the relaxation due to the local field (with a linewidth $\Delta_2\gtrsim0.15$ MHz) is suppressed by the motional averaging, namely, 
\begin{equation}
G_z(t)=f_1G_z^\mathrm{KT}(t;\Delta_1,\nu_1,B_\mathrm{LF})e^{-\lambda t}+f_2G^{\rm KT}_z(t;\Delta_2,\nu_2),\label{Eq:ana2}
\end{equation}
with $G^{\rm KT}_z(t;\Delta_2,\nu_2)\simeq1$ for $\nu_2\gg\Delta_2$.

Below $T_\gamma$, $\nu_1$ increases with decreasing $T$ in correlation with $f_2$, and tends to approach a constant value or decrease for $T\lesssim T_\alpha\simeq30$ K.  A similar correlation is observed between $f_2$ and $\lambda$; $\lambda$  gradually increases with decreasing $T$ to exhibit a small peak around $T_\gamma$, then decreases towards $T_\beta$, followed by an increase to reach another maximum near $T_\alpha$.  Here, we point out that, as shown in Fig.~\ref{fig:f2_Ne_lmd}(b), the complicated $T$ dependence of $f_2$ and the increase of $\lambda$ occur in correlation with the decrease of $N_\mathrm{e}\mu_{\rm e}$  (with $\mu_{\rm e}$ being the mobility) of the sample used for $\mu$SR measurements.  More specifically, $N_\mathrm{e}$ exhibits gradual decrease around $T_\gamma$ with decreasing $T$, followed by a sharp decrease below $T_\beta$ and an increase below $T_\alpha$  (see Fig.~S2 in SM \cite{sm}).  As shown in Fig.~\ref{fig:f2_Ne_lmd}(d), $\mu_{\rm e}$ decreases with decreasing $T$ below $\sim$120~K, reaching a minimum around $T_\beta$, and then increases to a maximum at $T_\alpha$.  The behavior of $\mu_{\rm e}$ for $T>T_\gamma$ is qualitatively in line with previous reports~\cite{Parisini:16,Kabilova:19}.  This implies that the electronic state of Mu is highly sensitive to the quality of samples, which is also supported by the fact that the $T$ dependence of $f_2$ (and $f_1$) in powder sample differs significantly from that in single crystal (see Fig.~S5(b) in SM \cite{sm}).

 \section{DISCUSSION}\label{Dcn}
\begin{figure}[t]
  \centering
	\includegraphics[width=0.95\linewidth,clip]{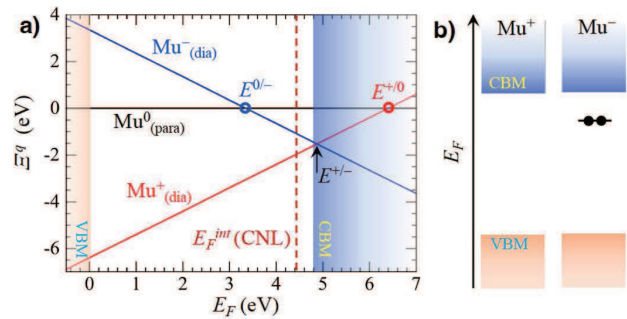}
	\caption{a) The formation energy ($\Xi^q$) of interstitial Mu (Mu$^q_{\rm i}$, $q=0,\pm$) vs the Fermi level ($E_F$) in \gao\ inferred from DFT calculations for H$^q$ \cite{DFT_Li:14}.  The donor/acceptor levels are determined as cross points between $\Xi^\pm(E_F)$ and $\Xi^0(E_F)$. The dashed line shows the intrinsic charge neutral level ($E_F^{\rm int}$) obtained from the DFT calculation. b) Schematic band diagrams for the electron energy associated with the donor/acceptor levels in a). }
	\label{FE}
\end{figure}

In the ambipolarity model, the electronic state of the interstitial Mu is determined by where the donor and acceptor levels associated with H, predicted from the Fermi energy ($E_F$) dependence of defect formation energy ($\Xi^q$) for H$^q$  ($q=\pm,0$), is located in the energy band structure of the host material.  Figure~\ref{FE}(a) shows $\Xi^q(E_F)$ for each valence state of H obtained by previous DFT calculations \cite{DFT_Li:14}, where H is substituted with Mu. The relationship between the donor/acceptor levels and band structure predicts that Mu can take two different diamagnetic states: a donor-like state in which it releases an electron into the conduction band to become Mu$^+$, and a hydride (Mu$^-$) state, which is qualitatively consistent with the present observations. Interestingly, the $E^{+/-}$ level, which determines the charge state of H at thermal equilibrium, is in the conduction band as is the donor level, so the behavior of H is expected to be the same as that of the donor-like Mu.

\begin{table}[b]
	\centering
	\caption{Comparison of the linewidth ($\Delta$) between those at the H sites obtained from structural relaxation calculations and the experimental value. $\Delta_{\rm sc}$, and $\Delta_\mathrm{pwdr}$ for single crystalline (sc) and powder (pwdr) samples, respectively. $d_\mathrm{OH}$ is the distance from H to the nearest oxygen (see also Fig.~\ref{fig:Rlx_str}).}
  \renewcommand\arraystretch{1.2}
  \tabcolsep=1.5mm
  \begin{tabular}{cccc}
		\hline\hline
    Site & $d_\mathrm{OH}$ [nm]& $\Delta_{\rm sc}$ (MHz) & $\Delta_\mathrm{pwdr}$ (MHz) \\ \hline
    H$_\mathrm{I}$   & 0.0966 & 0.136 &  0.137 \\
		H$_\mathrm{\rtwo}$  & 0.1036 & 0.150 &  0.153 \\
		H$_\mathrm{\rthree}$ & 0.1003 & 0.161 &  0.156 \\
		H$^-$ &  -- & 0.345 & 0.299 \\
    \hline
   $\Delta_1$ (300~K)& & 0.140(1) &  0.136(1)\\
    \hline \hline
 	\end{tabular}
	\label{tbl:delta_sim}
\end{table}
\par
\begin{figure}[t]
  \centering
	\includegraphics[width=0.7\linewidth,clip]{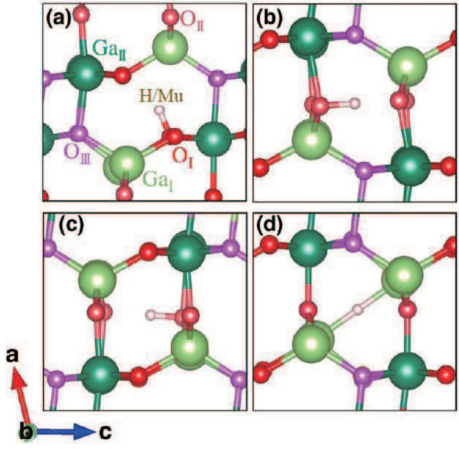}
	\caption{Local structures H-related defects corresponding to (a) H$_\mathrm{I}$, (b) H$_\mathrm{\rtwo}$, (c) H$_\mathrm{\rthree}$ and (d) H$^-$ state.  H$_\mathrm{I-III}$ are bonded to O, while H$^-$ forms bonding with two Ga$^{3+}$ ions \cite{DFT_Li:14}.   The crystal structures were displayed using VESTA~\cite{Vesta}.}
	\label{fig:Rlx_str}
\end{figure}

To determine the local defect structure of the Mu$_1$ state, $\Delta_1$ was compared with those calculated for the candidate sites, and the results are summarized in Table ~\ref{tbl:delta_sim}.  The value for the H$_\mathrm{I}$ site bonded to the three-coordinated O$_\mathrm{I}$ [see Fig.~\ref{fig:Rlx_str}(a)] is in good agreement with $\Delta_1$ at 300 K, indicating that H$_{\rm I}^+\approx{\rm Mu}_1^+$ in the dilute limit ($\sim$10$^5$~cm$^{-3}$). This confirms the earlier point that, aside from not knowing the exact location of $E^{+/-}$, H$_\mathrm{I}$ can serve as donor~\cite{DFT_Varley:11}.  The existence of H in the H$_\mathrm{I}$ structure hss been also inferred from the infrared spectroscopy of hydrogenated \gao~\cite{Qin:19_IR2}.
 The possibility that Mu$_1$ corresponds to H trapped in Ga and O vacancies is excluded by the comparison of $\Delta_1$ with those for Mu in these vacancies (shown in Table~S1 in SM \cite{sm}).

As seen in Fig.~\ref{fig:asyTdep_par}(c), $\Delta_1$ exhibits a slight increase with decreasing $T$, approaching the value at the H$_{\rm II}$/H$_{\rm III}$ sites [see Fig.~\ref{fig:Rlx_str}(b), (c)]. This can be interpreted as reflecting the fact that the initial population of relaxed-excited states immediately after muon implantation is a random sampling of available metastable sites. Considering that the difference in the formation energy estimated by DFT calculations is small among these sites (e.g., $\sim$30 meV between H$_{\rm I}$ and H$_{\rm II}$), the decrease in $\Delta_1$  with increasing $T$ above $T_\gamma$ suggests that the Mu site distribution approaches the lowest energy state by the annealing process. A similar $T$ dependence of $\Delta$ exhibited by donor-like Mu has been reported for InGaZnO$_4$ in which multiple H sites are available \cite{Kojima:19_IGZO}.

With Mu$_1$ identified as being in the donor-like state, the paired Mu$_2$ component is presumed to be in the acceptor-like diamagnetic state (Mu$_2^-$) from the ambipolarity model.  According to DFT calculations, the corresponding H$^-$ state is bonded to the two nn Ga ions, as shown in Fig.~\ref{fig:Rlx_str}(d) \cite{DFT_Li:14}. The $E^{0/-}$ level is near the conduction band, and it can still accommodate two electrons as it is located below the charge neutral level ($E_F^{\rm int}$, see Fig.~\ref{FE}).   This means that it is possible for Mu$_2$ to exchange electrons with the conduction band. Moreover, the conduction band has a relatively large dispersion (effective mass $\sim$0.28$m_e$) around the $\Gamma$ point \cite{Peelaers:15}, allowing fast migration of electrons.  
\par
While Mu$_2^-$ is unlikely to exhibit fast diffusion by itself due to the bonding to cations, there are many reported examples of fast diffusive motion of acceptor-like Mu$^0$ in semiconductors and alkali halides~\cite{Patterson:88,Kadono:94_GeAs,Gxawu:05_C,Chow:2000_Si}. 
Since $N_{\rm e}\mu_{\rm e}$ and $f_2$ show almost the same $T$ dependence for $T>T_\beta$, it is inferred that the Mu$_2$ state diffuses rapidly for $T>T_\gamma$ through the neutral state temporally attained by the charge exchange reaction Mu$_2^-\rightleftarrows{\rm Mu}_2^0+e^-$.  The reaction rate is given by  
\begin{equation}
r_\mathrm{ex}=\frac{1}{\tau_{-0}+\tau_{0-}} =\frac{r_{-0}r_{0-}}{r_{-0}+r_{0-}},\label{rex}
\end{equation}
 where $r_{-0}$ ($=1/\tau_{-0}$) is the ionization probability ($\propto e^{-\varDelta E/k_BT}$, with $\varDelta E\sim E_c-E^{0/-}$ and $E_c$ the CBM energy) and $r_{0-}$ ($=1/\tau_{0-}$) is the capture rate of the free carriers ($\propto \sigma_{\rm c}N_\mathrm{e}\mu_{\rm e}V_r$, with $\sigma_{\rm c}$, $\mu_{\rm e}$, and $V_r$ being the cross section, carrier mobility, and electric field exerted from the Mu-Ga complex state, respectively). Note that $\lambda$ due to the charge exchange is also quenched by the motional averaging when $r_\mathrm{ex}$ is much greater than the hyperfine parameter. Such motional effects have also been observed in SiO$_2$ and are thought to be responsible for the disappearance of anisotropy in the hyperfine interaction of Mu$^0$ with increasing $T$~\cite{Chow:2000_Si}.
The existence of the non-relaxing Mu component has also been reported for variety of materials including NaAlH$_4$, LaScSiH$_x$, and FeS$_2$~\cite{Kadono:08_NaAlH4,Okabe:18_FeS2,Hiraishi:21_LaScSi}, implying the ubiquitous nature of acceptor-like Mu states.
\par
It is natural to assume that Mu$_2$ diffuses along the $\langle010\rangle$ direction because \gao\ has an open channel structure along the $\langle010\rangle$ direction,
In fact, secondary ion mass spectrometry analysis of $^2$H implanted \gao~\cite{Vilde:20_Hdiffusion} revealed that $^2$H diffuses more easily in the $\langle010\rangle$ direction than perpendicular to the $\langle-201\rangle$ surface.
It has also been reported for rutile TiO$_2$ that hydrogen diffusion is more likely to occur in open channels along the $c$-axis than in the $a$-$b$ plane~\cite{Bates:79_TiO2}.
\par
There are three different open channels in the $\langle010\rangle$ direction in \gao.
The potentials associated with diffusion in each channel were investigated by total energy calculations (see Sect.~III in SM~\cite{sm} for details), and it was found that hydrogen tends to diffuse in the same channel as the Mu$_1$ site. Since the two charged states Mu$_1^+$ and Mu$_2^-$ accompany mutually different lattice distortions, it is assumed that the potential barriers caused by these distortions allow Mu$_2$ to diffuse while maintaining its state. However, because the adopted calculations do not take structural relaxation into account, it is a future task to evaluate using the nudged-elastic band (NEB) method.
\par
The details of what causes the decrease in $f_2$ (increase in $f_1$) and recovery below $T_\gamma$, which is apparently correlated with the onset of slow dynamics for Mu$_1$ (as inferred from the $T$ dependence of $\Delta_1$, $\nu_1$, and $\lambda$), are currently unknown. Nevertheless, when analyzing the time spectra in this $T$ range using Eq.~(\ref{Eq:ana2}) with $\Delta_1$ and $\Delta_2$ as free parameters, $\Delta_i$  converge to mutually close values ($\sim$0.13--0.14 MHz). A curve fit performed with $\Delta_1$ fixed at 0.136 MHz and $\Delta_2$ at 0.345 MHz (corresponding to the stationary hydride state) and the other parameters free fails to reproduce the data, as indicated by a large reduced chi-square (see Fig.~S6 in SM). Thus, the decrease in $f_2$ suggests that a transition from Mu$_2$ to Mu$_1$ occurs in the relevant $T$ range. One possible scenario is that  Mu$_2^-$ can be ionized by thermal excitation and transitions to Mu$_1^+$ for $T<T_\beta$, but the ionization is prevented by increasing $N_{\rm e}\mu_{\rm e}$ for $T>T_\beta$ [i.e., $r_{0-}\gg r_{-0}$ in Eq.~(\ref{rex})], allowing it to continue to exist as a metastable state again above $T_\gamma$. 
\par
Finally, we comment on the relationship between our experimental results and the prior $\mu$SR studies on \gao. In Ref.~[\onlinecite{musr_King:10}], TF-$\mu$SR measurements are reported for a powder sample from a commercial vendor, and it is observed that about 10\% of the signal exhibits slight increase of relaxation rate from 0.08 MHz to 0.12 MHz below 50--100 K. Ref.~[\onlinecite{musr_Celebi:12}] reports results for a single-crystal sample (orientation unknown) and observes a 4--6\% decrease in the initial asymmetry below 50--100 K in both ZF and TF-$\mu$SR measurements and an increase in $\Delta$ of the component described by $G_z^{\rm KT}(t)$ from 0.10--0.11 MHz to 0.15 MHz in the ZF-$\mu$SR spectra.  Although detailed comparisons with our results are not possible because no time spectra are available for either case, these changes appear to correspond to the increase in $f_1$ at $T<T_\gamma$ that we observed in both the powder and single-crystalline samples [see Fig.~S5(b) in SM \cite{sm}]. Meanwhile, these reports are silent about the component corresponding to Mu$_2$. We stress that the existence and origin of Mu$_2$ are clearly identified for the first time in this study with the help of the ambipolarity model.

\begin{acknowledgments}
This work was supported by the MEXT Elements Strategy Initiative to Form Core Research Center for Electron Materials (Grant No.~JPMXP0112101001) and JSPS KAKENHI (Grant No.~19K15033).
The $\mu$SR experiments were conducted under user programs (Proposal No.~2019MS02) at the Materials and Life Science Experimental Facility of the J-PARC.
We also acknowledge the Neutron Science and Technology Center, CROSS for the use of PPMS in their user laboratories.
\end{acknowledgments}

\input{Ga2O3.bbl} 

\newpage
{\ }

\newpage
\setcounter{section}{0}
\setcounter{figure}{0}
\setcounter{table}{0}
\setcounter{equation}{0}
\renewcommand{\thefigure}{S\arabic{figure}}
\renewcommand{\thetable}{S\arabic{table}}
\renewcommand{\theequation}{S\arabic{equation}}

\noindent
\textbf{Supplemental Material: Local electronic structure of dilute hydrogen in $\beta$-Ga$_2$O$_3$ probed by muons 
}
\newline 
\begin{center}
{M. Hiraishi {\it et al.}}
\end{center}

\section{Properties of Single Crystalline $\beta$-${\bf Ga_2O_3}$}
\subsection{Thermal Desorption Spectrometry (TDS)}
\begin{figure}[htbp]
  \centering
  \includegraphics[width=0.4\textwidth]{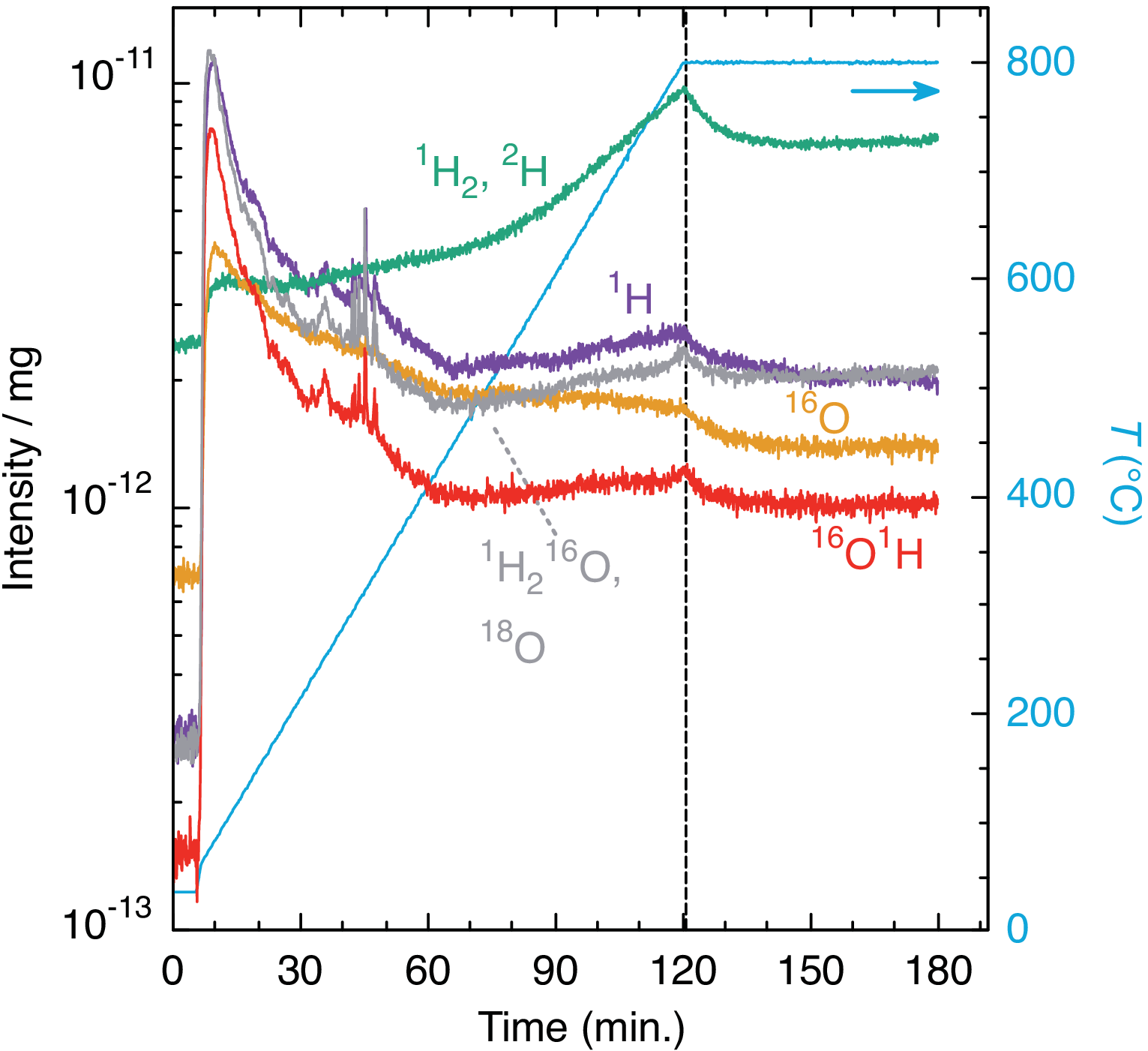}
  \vspace{-2mm}
  \caption{TDS spectra of the $\beta$-Ga$_2$O$_3$ sample used for $\mu$SR measurements. Purple, green, yellow, red, and gray are signals for $m/z=$1, 2, 16, 17, and 18, respectively ($m$: mass number, $z$: ion valence). Temperature history is shown on the right vertical axis.}
  \label{FigS_TDS}
\end{figure}

Figure~\ref{FigS_TDS} shows the TDS results for the $\langle001\rangle$ sample, which was linearly heated from room temperature to 800$^\circ$C and held at 800$^\circ$C for 1 hour.
Temperature is shown on the right axis and the signal intensities obtained by the quadrupole mass spectrometer (QMS) are shown on the left vertical axis.
The analysis revealed that the $\langle001\rangle$ sample contains 3.5$\times10^{18}\mathrm{cm}^{-3}$ of hydrogen ($m/z=2$, with $m$ being the mass number, $z$ the ion valence).
Considering that the desorption process is diffusion-limited for crystalline specimens, this value may correspond to the lower limit.
The signal intensity for $m/z=18$ ($^1$H$_2^{16}$O, $^{18}$O) is greater than that for $m/z=16$ ($^{16}$O) multiplied by 0.002 (natural abundance of $^{18}$O), indicating that $m/z=18$ is derived from water (H$_2$O).
The similar temperature dependence of $m/z=1$ and 17 to that of 18 strongly suggests that $m/z=1, 17$ are also water-related signals.
The steep increase below 100$^\circ$C ($\sim$20~min.) is probably due to residual moisture on the sample surface, chamber, and so on.

\subsection{Carrier concentration}
\begin{figure}
  \centering
  \includegraphics[width=0.7\columnwidth]{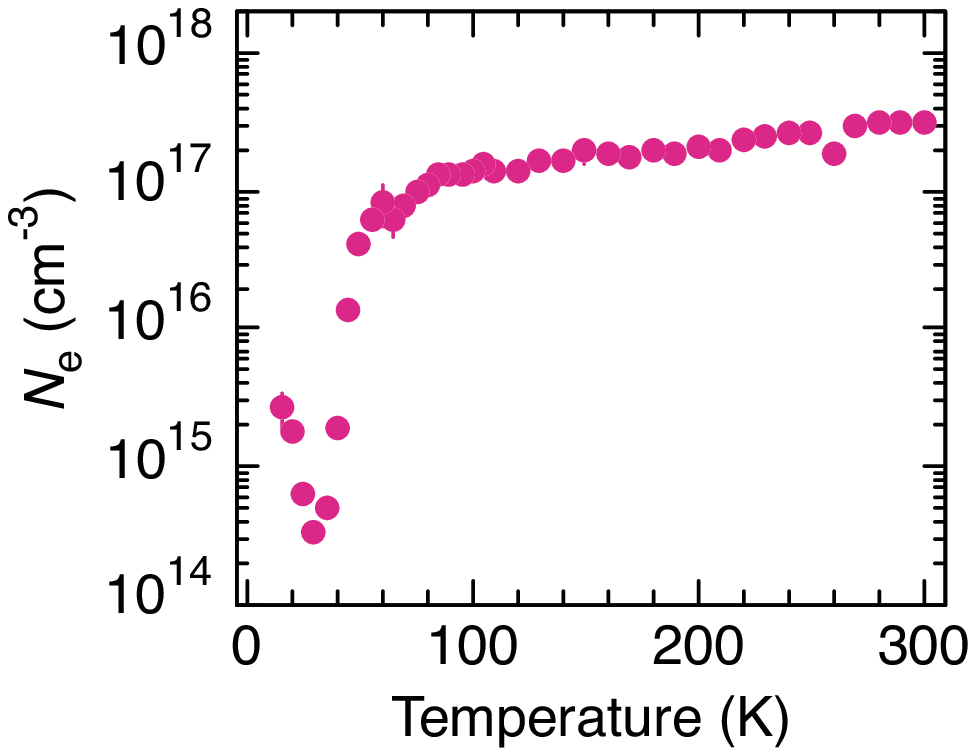}
  \vspace{-2mm}
  \caption{Temperature dependence of the electron carrier concentration in the single crystalline sample of $\beta$-Ga$_2$O$_3$.}
  \label{FigS_Ne}
\end{figure}

The temperature dependence of carrier concentration obtained from Hall coefficient measurements for the $\beta$-Ga$_2$O$_3$ sample (with $\langle001\rangle$ orientation) used for $\mu$SR measurements is shown in Fig.~\ref{FigS_Ne}. The sign of the Hall coefficient indicates that majority carriers are electrons. 

\section{Result of DFT calculations}
\begin{figure*}[t]
  \centering
  \includegraphics[width=0.7\linewidth]{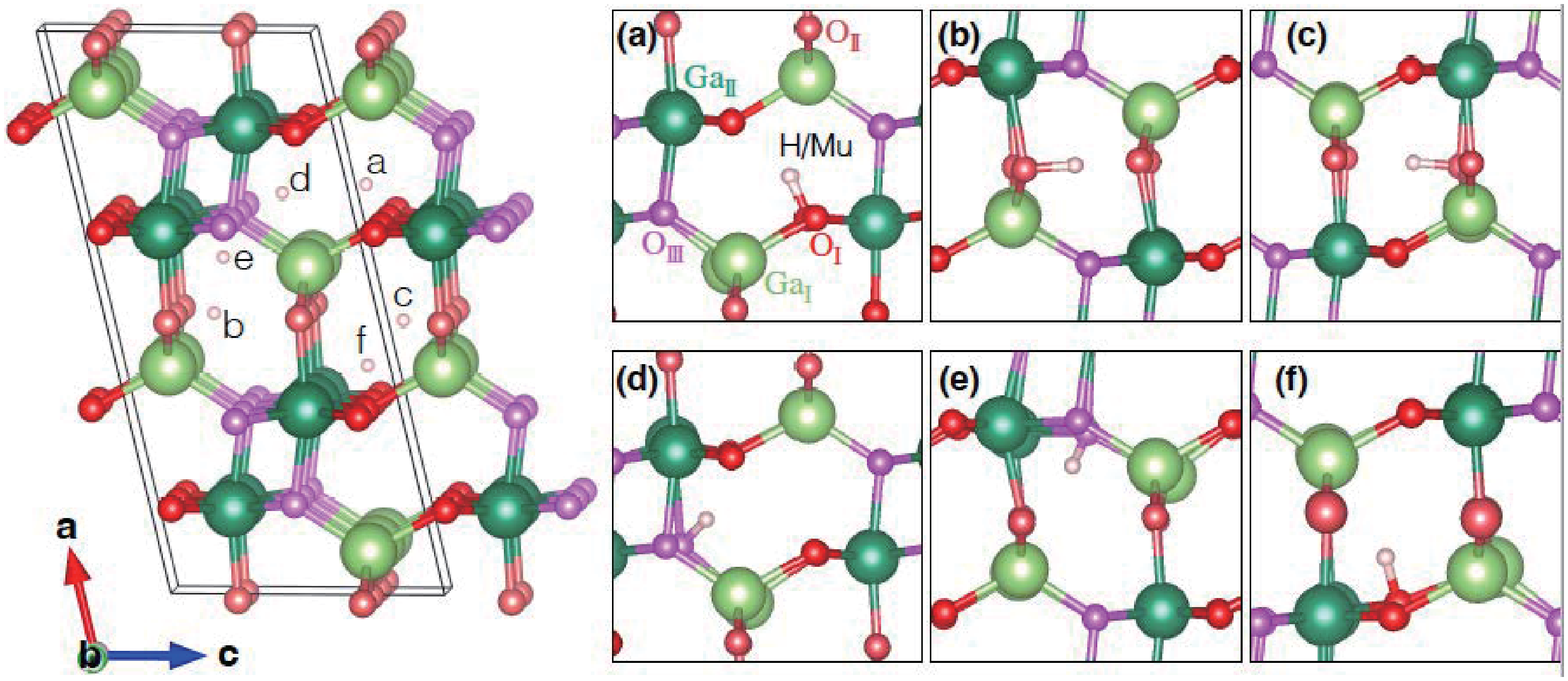}
  \caption{Left: Crystal structure of $\beta$-Ga$_2$O$_3$. Isolated pink balls indicate the initial positions of hydrogen in the structural relaxation calculations shown in (a)-(f), respectively corresponding to H$_\mathrm{I}$, H$_\mathrm{\rtwo}$, H$_\mathrm{\rthree}$, H$_\mathrm{\rfour}$, H$_\mathrm{V}$, and H$_\mathrm{\rsix}$ in the Table~\ref{tblS:delta_sim}. The crystal structures were drawn using VESTA~\cite{Vesta}. The nuclear dipolar linewidth $\Delta$ calculated for each structure is shown in Table~\ref{tblS:delta_sim}.}
  \label{FigS_RlxStr}
\end{figure*}

The relaxed structures obtained by DFT calculations are shown in Fig.~\ref{FigS_RlxStr}, and the simulated muon spin relaxation rates $\Delta$ in the Kubo-Toyabe function for each structure are shown in Table~\ref{tblS:delta_sim}.

\begin{table}[h]
	\centering
	\caption{The nuclear dipolar linewidth $\Delta$ for the various local structures involving interstitial H obtained from structural relaxation calculations. $\bigstar$ indicate the results of VASP:HSE06 calculations \cite{VASP,HSE06}, while others were those by OpenMX:GGA-PBE~\cite{OpenMX_Ozaki:03}. $d_\mathrm{OH}$ denotes the distance between H and the nearest O. $\Delta_{\langle001\rangle}$ and $\Delta_\mathrm{pwdr}$ are the linewidths for single crystalline and powder samples, respectively. }
  \vspace{2mm}
  \renewcommand\arraystretch{1.2}
  \tabcolsep=2mm
  \begin{tabular}{lcccll}
		\hline\hline
    \multirow{2}{*}{Type} &   \multirow{2}{*}{Config.} &   \multirow{2}{*}{nn oxygen} & $d_\mathrm{OH}$  & $\Delta_{\langle001\rangle}$ & $\Delta_\mathrm{pwdr}$ \\ 
     & & & (nm) & (MHz) & (MHz) \\
    \hline
    Interstitial$^\bigstar$ & H$_\mathrm{I}$            & O$_\mathrm{I}$       & 0.0966 & 0.136 &  0.137 \\
		Interstitial$^\bigstar$ & H$_\mathrm{\rtwo}$        & O$_\mathrm{\rtwo}$   & 0.1036 & 0.150 & 0.153 \\
		Interstitial$^\bigstar$ & H$_\mathrm{\rthree}$      & O$_\mathrm{\rtwo}$   & 0.1003 & 0.161 & 0.158 \\
		Interstitial & H$_\mathrm{\rfour}$       & O$_\mathrm{\rthree}$ & 0.1024 & 0.157 & 0.165 \\
		Interstitial	& H$_\mathrm{V}$            & O$_\mathrm{\rthree}$ & 0.0996 & 0.158 & 0.170 \\
		Interstitial	& H$_\mathrm{\rsix}$        & O$_\mathrm{I}$       & 0.0989 & 0.175 & 0.162 \\
    Vacancy                & V$_\mathrm{Ga_I}$-H       & O$_\mathrm{I}$       & 0.0983 & 0.106 & 0.117 \\
    Vacancy                & V$_\mathrm{Ga_{\rtwo}}$-H  & O$_\mathrm{I}$       & 0.0988 & 0.108 & 0.115 \\
		Vacancy                & V$_\mathrm{O_I}$-H        & ---                 & ---    & 0.286 & 0.250 \\
		Vacancy                & V$_\mathrm{O_{\rtwo}}$-H   & ---                 & ---    & 0.168 & 0.235 \\
		Vacancy                & V$_\mathrm{O_{\rthree}}$-H & ---                 & ---    & 0.229 & 0.238 \\
    \hline \hline
 	\end{tabular}
	\label{tblS:delta_sim}
\end{table}

\section{$\langle010\rangle$ Axial diffusion: Evaluation based on total energy.}
\begin{figure}[h]
  \centering
  \includegraphics[width=\linewidth]{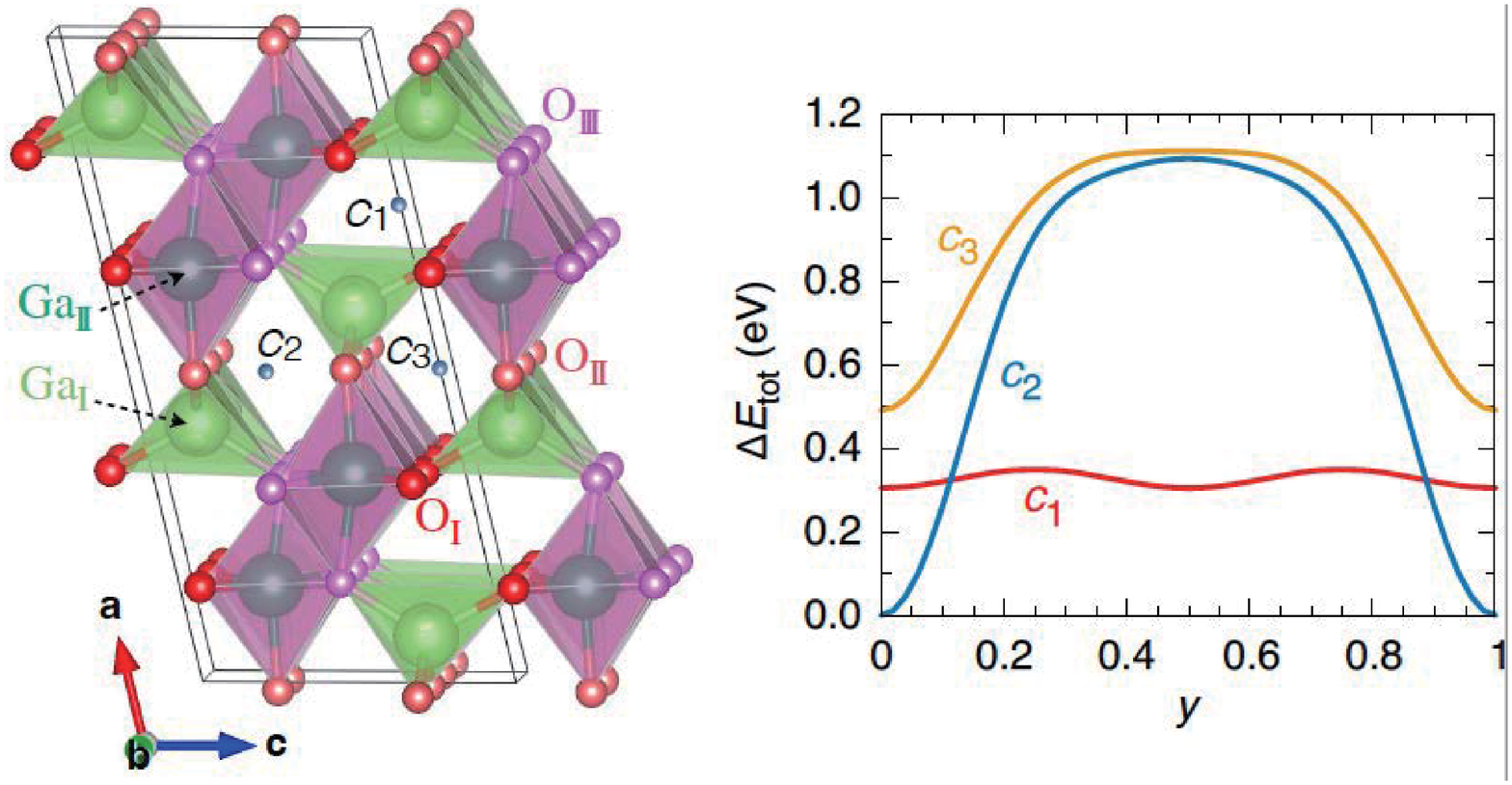}
  \caption{Hydrogen position on the $\langle010\rangle$ axis dependence of the total energy calculated at the center $c_i$ $(i=1,2,3)$ of each channel, shown in left panel, using the $1\times2\times1$ structure.}
  \label{FigS_010_dE}
\end{figure}

To investigate the Mu$_2$ diffusion path, $\langle010\rangle$ axis position $y$ dependence of the total energy $E_\mathrm{tot}$ was calculated in a $1\times2\times1$ superlattice.
Figure~\ref{FigS_010_dE} shows the variation of the total energy,
\begin{align}
\Delta E_\mathrm{tot}\equiv E_\mathrm{tot}({\bm r})-E_\mathrm{min},\nonumber
\end{align}
where $\bm{r}$ is the position vector of channel $c_i$ $(i = 1,2,3)$ and $E_\mathrm{min}$ is the global minimum of $E_\mathrm{tot}(\bm{r})$.
Here, ${\bm r}$ for $c_1=(0.25, y, 0)$, $c_2=(0.5, y, 0.5)$ and $c_3=(0.5, y, 0)$, respectively.
The $\Delta E_\mathrm{tot}$ of $c_2$ and $c_3$ show large changes of more than 0.5~eV, whereas the $\Delta E_\mathrm{tot}$ of $c_1$ is almost independent of $y$ and the difference is only 0.04~eV, suggesting that $c_1$ is a the possible diffusion path for Mu$_2$.

\section{Supplemental $\mu$SR Results}
\subsection{$\mu$SR results on powder Sample}
\begin{figure}[h]
  \centering
  \includegraphics[width=\columnwidth]{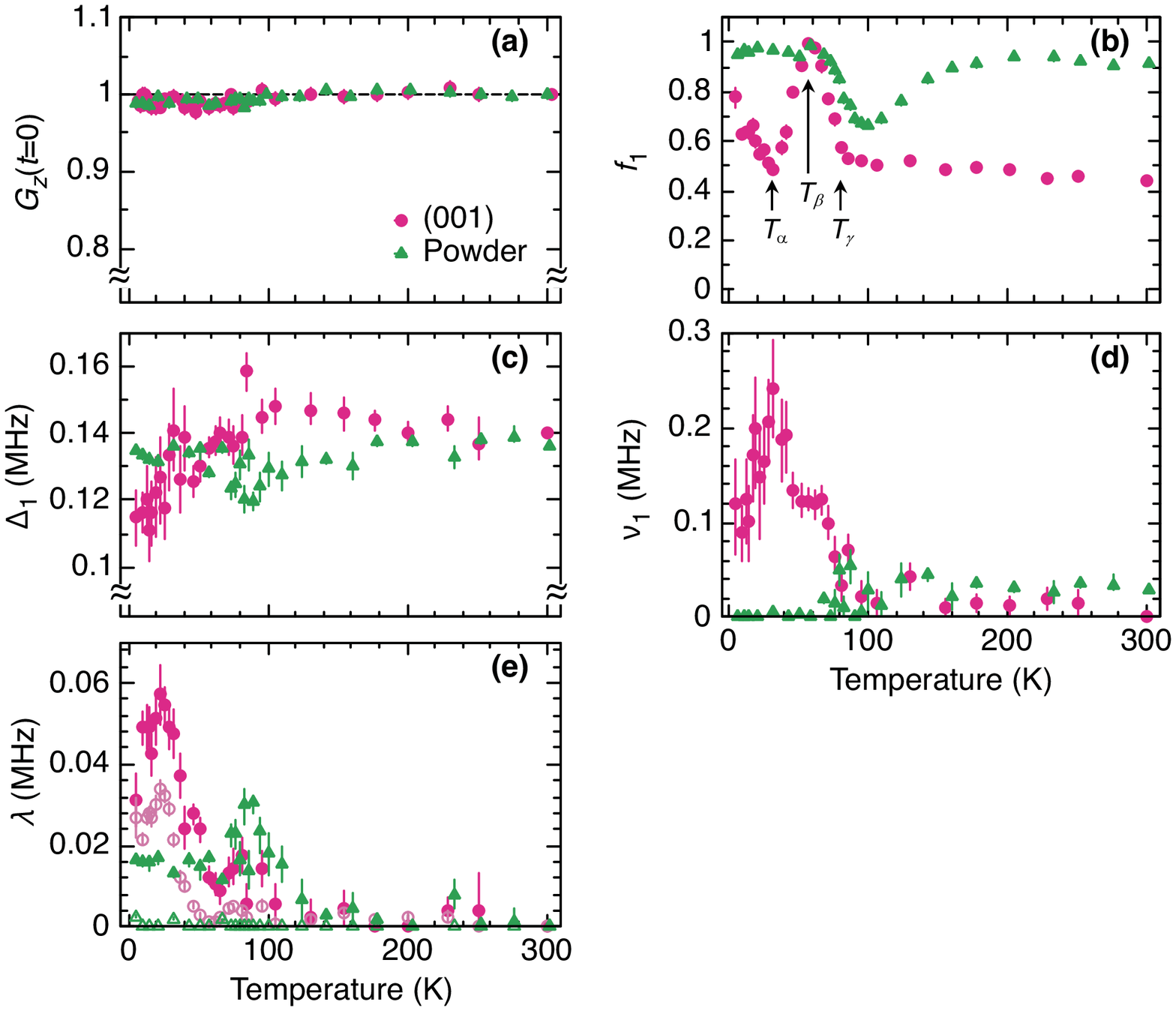}
  \vspace{-2mm}
  \caption{Temperature dependence of (a) $G_z(0)$ determined by TF-$\mu$SR, (b) the fractional yield $f_1$, (c) the Kubo-Toyabe linewidth $\Delta_1$, (d) the fluctuation rate $\nu_1$, and (e) the exponential relaxation rate $\lambda$ for the Mu$_1$ component. The result for the single crystalline sample is quoted from the main text for comparison.}
  \label{FigS_params}
\end{figure}

$\mu$SR measurements and data analysis for the powder sample were performed as for the single crystal.
The temperature dependence of each parameter obtained by curve fitting of the $\mu$SR time spectra is shown in Fig.~\ref{FigS_params}. For comparison, the results for the single-crystal sample presented in the main text are also shown. 
As can be seen in Fig.~\ref{FigS_params}(a), the most significant difference is that the yield $f_1$ of the Mu$_1$ component in the powder sample is greatly increased in place of the Mu$_2$ component. In addition, it is observed  for the powder sanple that $\nu_1$ (the fluctuation rate of $\Delta_1$) is almost zero (static) for $T\lesssim T_\gamma\simeq80$ K, whereas it takes on finite values for $T\gtrsim T_\gamma$, suggesting slow diffusive motion of the Mu$_1$ component. These behaviors are in remarkable contrast to the single-crystal sample.  On the other hand, the increase in $f_1$ (decrease in $f_2$) from $T_\beta$ ($\simeq55$ K) to $T_\gamma$ is common. From these observations, it may be interpreted that for $T<T_\beta$, the component corresponding to Mu$_2$ in the powder sample is bounded by some defects, etc., and that a part of it is detrapped and begins to exhibit fast diffusion  for $T_\beta<T<T_\gamma$, which is gradually suppressed above $T_\gamma$. 

\subsection{Comparison of curve fits for the data around $T_\beta$}
To test the possibility that quasi-static Mu$_1$ and Mu$_2$ coexist in the $\mu$SR spectrum near the temperature at which $f_2$ decreases significantly, we performed global curve fits for the ZF and LF-$\mu$SR spectra with the linewidth $\Delta_1$ and $\Delta_2$ as free parameters in each state, as well as for those fixed to the values expected for the corresponding sites ($\Delta_1 = 0.136$ MHz and $\Delta_2=0.345$ MHz). The results of the respective analyses for the spectra at 57 K are shown below in Fig.~\ref{FigS_57K}(a) and (b). The fits with $\Delta$ fixed to the expected values at the two sites show large residual chi-squares, clearly indicating that the data are not well reproduced.

\begin{figure}[h]
  \centering
  \includegraphics[width=\columnwidth]{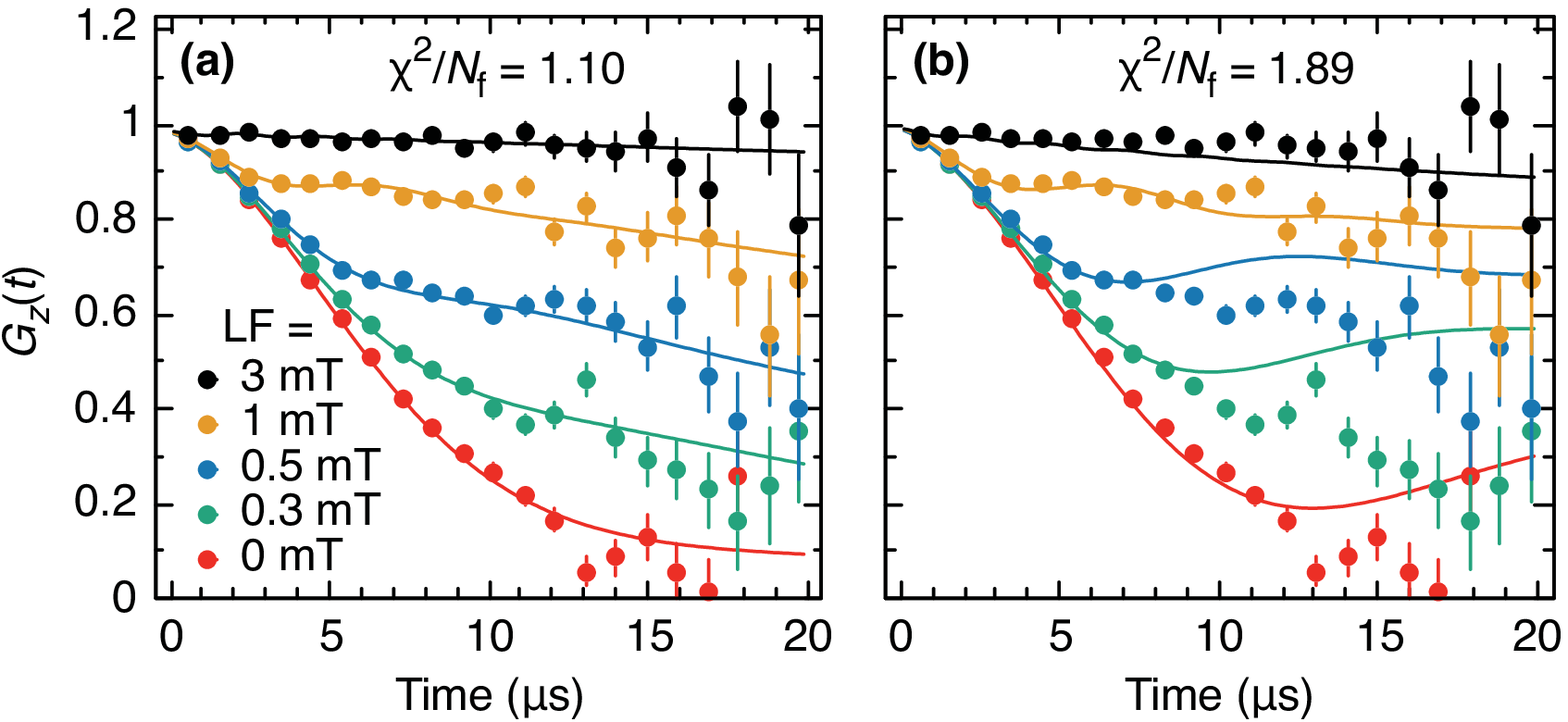}
  \vspace{-2mm}
  \caption{Analyzed $\mu$SR spectra at 57 K (global fit of the entire ZF/LF data). (a) with $\Delta_1$ and $\Delta_2$ as free parameters, (b) fixed to $\Delta_1 = 0.136$ MHz and $\Delta_2=0.345$ MHz.}
  \label{FigS_57K}
\end{figure}

\end{document}

%% file: Ga2O3.bbl
%